\documentclass[apjl]{emulateapj}
\usepackage{amsmath}

\shorttitle{Lorentz of GRBs}
\shortauthors{Sonbas et al.}

\begin{document}


\title{Gamma-Ray Bursts: Temporal Scales and the Bulk Lorentz Factor}


\author{E. Sonbas\altaffilmark{1,2}, G. A. MacLachlan\altaffilmark{2}, K.S. Dhuga\altaffilmark{2}, P. Veres\altaffilmark{2}, A. Shenoy\altaffilmark{2}, and T.N. Ukwatta\altaffilmark{3} }
\affil{$^1$University of Adiyaman, Department of Physics, 02040 Adiyaman, Turkey}
\affil{$^2$Department of Physics, The George Washington University, Washington, DC 20052, USA}
\affil{$^3$Department of Physics and Astronomy, Michigan State University, East Lansing, MI 48824, USA}
\email{edasonbas@yahoo.com}



\begin{abstract}
\hspace{-0.51cm} For a sample of Swift and Fermi GRBs, we show that the minimum variability timescale and the spectral lag of the prompt emission is related to the bulk Lorentz factor in a complex manner: For small $\Gamma$'s, the variability timescale exhibits a shallow (plateau) region. For large $\Gamma$'s, the variability timescale declines steeply as a function of $\Gamma$ ($\delta T\propto\Gamma^{-4.05\pm0.64}$). Evidence is also presented for an intriguing correlation between the peak times, t$_p$, of the afterglow emission and the prompt emission variability timescale.       
\end{abstract}



\section{Introduction}

\hspace{-0.53cm} For a majority of Gamma-Ray Bursts (GRBs), the emitted jet is highly relativistic, with a bulk Lorentz factor of a few hundred and produces, through internal/external shell collisions, very diverse and highly variable lightcurves. The observed prompt emission and the early afterglow provide important clues about the properties of the GRB central engine and the associated radiation mechanisms. To further our understanding of the prompt emission, the afterglow, and the possible connections to the activity of the central engine and the propagation phase of the jet, it is important to establish the nature of the link between key factors such as the bulk Lorentz factor and temporal properties such as the variability timescale and spectral lags.\\
\\
Liang et al. (2010) report strong mutual correlations among various timescales including the peak time (t$_p$) for afterglow optical and X-ray light curves. Assuming the peaks in the afterglow are indicative of the deceleration of the jet in a constant medium, they extract the initial bulk Lorentz factors ($\Gamma_0$) for a sample of GRBs. They also note the existence of a tight correlation between the extracted Lorentz factors and E$_{iso}$. Ghirlanda et al. (2012) also determine the bulk Lorentz factors using afterglow peak times. They assume different density profiles (i.e., homogenous and wind) for the environment medium in which the jet propagates, and find the correlations: E$_{iso}$ and L$_{iso}$$\propto$ $\Gamma_0^2$ and E$_{peak}$$\propto$ $\Gamma_0$. Their results are consistent with the magnetically accelerated jet model simulations that are described by Komissarov, Vlahakis \& Koenigl (2010), Tchekhovskoy, McKinney \& Narayan (2009), Tchekhovskoy, Narayan \& McKinney (2010). Lu et al. (2012), estimated Lorentz factors using beaming-corrected jet luminosity from a neutrino-driven wind medium and extended the correlations between $\Gamma_0$ and E$_{\gamma,iso}$.\\  
\\
Temporal variability is also connected to the bulk Lorentz factor. A robust method for extracting such variability was recently described by MacLachlan et al. (2012, 2013), in which the authors used a technique based on wavelets and analyzed a sample of Fermi/GBM GRBs and showed that a variability (related to the minimum timescale (MTS) that separates white noise from red noise) of a few milliseconds is not uncommon for GRBs. For similar studies involving a large $Swift$ sample see Dolek et al. (2014) and Golkhou et al. (2014). MacLachlan et al. (2013) also demonstrated that there is a direct link between the shortest pulse structures as determined by the MTS and pulse-fit parameters such as rise times as seen in GRB prompt emission. Recently, Sonbas et al. (2013) applied the same technique to analyze X-ray flares and confirmed the validity of the relation between the MTS and pulse-fit parameters and extended the relation from the prompt emission to X-ray flares covering a temporal scale ranging over several orders of magnitude.\\
\\
In this paper, we explore the connection between the temporal variability timescale ($\delta T$), the peak times ($t_{p}$) for the optical emission, and the GRB bulk Lorentz factor ($\Gamma$). The paper is organized as follows: In section 2 we outline briefly the procedures used in the selection and processing of the GRB data, and the extraction of the minimum variability time scales. We also mention the methods employed in extracting the bulk Lorentz factors. In section 3 we present and discuss the correlations between the temporal variables and the Lorentz factors. Our summary and conclusions are presented in the last section. 
   
\section{Data and Methodology}

\hspace{-0.3cm}We have generated mask-weighted, background subtracted light curves by using $batgrbproduct$, $batmaskwtevt$, and $batbinevt$ tasks. These light curves were generated with a time binning of 200 $\mu$seconds and 100 $\mu$seconds in the four standard Swift/BAT energy bands, i.e. 15 - 25 keV, 25 - 50 keV, 50 -100 keV, 100 - 150 keV.\\ 
\\
For the Fermi sample we have extracted light curves for the GBM NaI detectors over the entire energy range (8 keV to 1 MeV). Typically, the brightest three NaI detectors were chosen for the extraction. Light curves for both long and short GRBs were extracted at a time binning of 200 $\mu$seconds.\\
\\
The time variability of the light curves was extracted by employing the wavelet technique of MacLachlan et. 2013. In this method the variability is related to the minimum time scale that separates the white noise and the red noise in a power density spectrum depicted by the dominant wavelet coefficients used to represent the sample lightcurves. For full details of the data and MTS extraction procedures, we refer the reader to MacLachlan et al. (2013). The spectral lags were extracted by using the technique described by Ukwatta et al (2012).\\
\\
A number of methods exist for extracting the bulk Lorentz factor: (1) One method relies on knowing the peak time of the early afterglow and relating that to the deceleration time of the external forward shock. This technique (Sari and Piran 1999) is used to extract the so-called 'initial' Lorentz factor which is approximated to be twice the Lorentz factor at the deceleration time, (2) The use of the 'compactness' condition (sometimes referred to as the 'opacity' argument) (Piran 1999), the requirement that bursts are optically thin to pair production. This approach (Lithwick and Sari 2001; Gupta and Zhang 2008; Abdo et al. 2009) yields a lower limit for the initial $\Gamma_0$, (3) The use of the approximation that the emission due to the external shock is not prominent during the prompt phase (Zou and Piran 2010), and finally, (4) Pe$^{\prime}$er et al (2007) describe a method based on the observation that a thermal component may play a significant role in the prompt emission in GRBs.\\
\\
We take the $\Gamma's$ from the samples of Lu et al (2012) and Ghirlanda et al. (2012). We assigned error bars of $10\%$ to the Ghirlanda values because the original data set came without uncertainties. We discarded a number of GRBs from the Lu et al. (2012) sample because only a limit is given for the Lorentz factors. We include a number of Fermi GRBs that have been recently analyzed by Ghisellini et al. (2010) and Kumar and Duran (2010). These GRBs are particularly interesting because they have significant prompt emission in the MeV and GeV energy region as well as late-time X-ray and optical emission, and have relatively high Lorentz factors compared to the Swift sample. We list in Table 1 the GRBs used in our analysis.\\ 

\section{Results and Discussion}

From kinematics alone (Fenimore et al. 1996, Dermer 1998, Salmonson 2000, Sultana et al. 2012), a connection between the time variability and the bulk Lorentz factor is expected. Regarding the temporal structures in GRBs, Kobayashi et al. (1997) and Wu et al. (2000) mention several timescales; in particular, they note the angular spreading time scale. This scale is the delay between the arrival times of the photons emitted at the line of sight and those emitted at some latitude along the side of the spreading region of the shell. As the flow is relativistic, the beaming effect of the radiating particles leads to emission from a narrow cone with an opening angle of $\sim$ 1/$\Gamma$. This produces a time delay of the order $t_{ang} \sim$ $\Gamma^{-2}$. If we assume this scale is related to the variability time scale, then we can expect a correlation between MTS and the bulk Lorentz factor. We explore such a possibility by plotting MTS versus the reported Lorentz factor, $\Gamma$: See Figure 1. The MTS is corrected for the 1/(1+z) time dilation factor, where z is the redshift for a given burst. The data exhibit a very interesting knee-like feature in the MTS-$\Gamma$ plots: The MTS shows a shallow (plateau) region for small $\Gamma$'s and then decreases rather sharply for high $\Gamma$'s. The knee or break in $\Gamma$ lies in the range 200 - 300. We have fitted the data with a broken power-law function (with the plateau region treated as a constant); the fit is depicted in the figure as a solid line. The best-fit for the sloped region for the Lu et al. (2012) sample is -3.84 $\pm$ 0.30 and the one for the Ghirlanda et al. (2012) sample (not shown) is -4.27 $\pm$ 0.64. The mean of the fitted break for the two data sets is $\Gamma = 225\pm30$ and beyond this $\Gamma$, the MTS declines steeply. \\
\\
The rapidly declining phase of MTS - $\Gamma$ can be qualitatively viewed as an opacity effect. One of the main lines of argument for associating GRBs with relativistic sources is the fact that we observe GeV photons (and non-thermal spectra), whereas in a non-relativistic source these photons would pair-produce and MeV-GeV range photons would not be observable. This argument requires Lorentz factors of the order of 100 or greater to reduce the pair creation optical depth so that high energy photons can escape.\\
\\
We take the optical depth to pair production from Lithwick \& Sari (2001) and express the Lorentz factor for $\tau_{\gamma\gamma}\approx 1$ (the transparency condition), and get:~$\Gamma\propto\delta T ^{-1/(2\beta+2)}$. The temporal variability $\delta T$ enters this expression as a characteristic timescale for the burst governing the radius of the emission, and we associate it with the MTS. The exponent $\beta$ is the photon index of the spectrum, assumed to be a simple power law. A characteristic value from observations can be the taken as the high-energy index of the Band function (Band et al. 1993) i.e., $\beta = 2$.  With this value we recover the limiting behavior MTS $\propto \Gamma^{-6}$ at high energies. We note this is the lower limit of $\Gamma$. In this interpretation the steep decline of the MTS with increasing Lorentz factor can be seen as a "bias" i.e., for very small variability times, one needs a high Lorentz factor to avoid pair creation and produce a GRB.\\
\\
To understand the plateau region of the MTS-$\Gamma$ plots, we follow the work of Gupta and Zhang (2008). These authors note that the relation between the Lorentz factor, $\Gamma$, the radius of the emission site, R, and the variability time scale, $\delta T$, need not follow the usual relation that is often used i.e., R $\propto$ $\Gamma^{2}c\delta T$/(1+z). They point out that while this relation is appropriate in the case of the internal shock model, it is not necessarily applicable for photospheric models nor models that invoke magnetic dissipation as the leading cause of the prompt emission. Gupta and Zhang (2008) develop an expression for the opacity to pair production without the internal shock assumption. Their result for R, the radius of the emission site, has the following form:
\begin{eqnarray}
\resizebox{.5\hsize}{!}{R$^{2}\propto$ A(E)$(\Gamma/(1+z))^{(2 - 2\beta)}$}
\end{eqnarray}
where A(E) carries terms relating to the luminosity distance, the flux, the Thompson cross section and a threshold energy related to pair production. If we assume that R is proportional to the variability time scale MTS, and that the index $\beta$ is in the range 1 - 2, then we can expect from the above expression a slope in the range -( 0 - 1) for a plot of MTS versus $\Gamma$ i.e., a very shallow dependence of MTS with $\Gamma$, which is what the data seem to indicate for low $\Gamma$'s. Indeed, Gupta and Zhang (2008) further show that for certain energy constraints i.e., when the threshold energy falls below the break energy but is still higher than the overall cutoff energy of the spectrum, the expression for the opacity (for $\beta$ = 1) is such that the radius R and the bulk Lorentz factor, $\Gamma$, become decoupled from each other. In this energy regime, $\Gamma$ is not directly related to R, implying no obvious connection between MTS and $\Gamma$.\\  
\\
In the curvature model, the spectral lag is expected to scale with the angular spreading time scale (t$_{ang}$), i.e., spectral lags should also exhibit a correlation with the bulk Lorentz factor. Indeed, Qin et al. (2004), Shen et al. (2005), and Lu et al. (2006), have examined the effect of the bulk Lorentz factor on spectral lags. Shen et al. (2005), show that spectral lag decreases with the Lorentz factor as lag $\propto$ $\Gamma^{-1}$. However, Lu et al. (2006), suggests that the spectral lag decreases much more rapidly as a function of the Lorentz factor i.e., as lag $\propto$ $\Gamma^{-\epsilon}$ where $\epsilon$ $>$ 2. Both models predict a negative correlation but with significantly different exponents. As seen in Figure 2, a negative correlation is seen for large $\Gamma's$. Large $\Gamma's$ imply small wavefronts and therefore smaller path lengths due to curvature, leading to smaller lags. Small $\Gamma's$ on the other hand mean large wavefronts and thus greater paths lengths. However, the degree of coherence is expected to decrease with the size of the wavefront, and with it, diminishing the variability at small timescales, thus producing little or no dependence of spectral lag with $\Gamma$.\\
\\
In the standard internal-external shock scenario the GRB is the result of internal shocks produced by a relativistic flow and the afterglow is produced via external shocks when the GRB flow interacts with the interstellar medium (ISM). According to Sari and Piran 1999, the reverse shock produced as a result of the impact can contain as much energy as the GRB itself but is at much lower temperature compared to the related forward shock and hence radiates at lower frequencies. These frequencies are typically in the optical regime. In order to ascertain if there is any correlation between a temporal scale for the GRB i.e., the prompt emission, and that for the optical emission in the afterglow, we plot in Figure 3, the peak times for the optical emission (taken from Liang et al. 2010) versus the MTS for the prompt emission. Although the error bars are large, a clear positive correlation between these temporal properties is seen in the plot. A simple power law leads to the following best-fit:
\begin{eqnarray}
\resizebox{.9\hsize}{!}{log~(MTS) = (1.04 $\pm$ 0.05) log~(t$_p$) - (2.45 $\pm$ 0.10)}
\end{eqnarray}
\hspace{-0.1cm}This new correlation, if further substantiated by additional and more precise data, is very intriguing because it suggests a direct link between temporal properties of the prompt emission and the afterglow optical emission. Using the simple internal-external shock scenario, Sari and Piran 1999 show that the evolution of the GRB can be broken into two main stages, an initial stage where the Lorentz factor ($\Gamma_0$) is constant and a latter stage, often referred to as the deceleration stage, is characterized by the evolution of the Lorentz factor ($\Gamma$) as $t^{-3/8}$. It is from this measured decay (of light curves in the optical emission) that the extracted Lorentz factor (for the afterglow) is related back to the initial Lorentz factor ($\Gamma_0$) for the GRB (see Sari and Piran 1999, Panaitescu and Kumar 2000; Meszaros 2006), and Liang et al 2010).\\
\\
The optical peak occurs at the deceleration time and in the constant ISM can be expressed as:  $t_p\approx 10 {\rm s}~ E_{53}^{1/3} n_0^{-1/3}\Gamma_{300}^{-8/3}$ for an adiabatic blast wave, and $t_p\propto\Gamma^{-7/3}$ in the radiative case. If we think of the MTS as a variability timescale which through causality arguments defines the size of the emitting region, we can write MTS$\propto$ R/$\Gamma^2$ where R is the size of the emitting region. By relating the two expressions through the Lorentz factor, we get MTS$\propto t_p^{3/4}$ for the adiabatic case and MTS$\propto t_p^{6/7}$ for the radiative case. The extracted exponent from the best fit (see Fig. 3) is 1.04 $\pm$ 0.05. While not conclusive, the MTS-t$_p$ fit seems to favor a coefficient closer to (6/7) and therefore suggestive of radiative dissipation of the blast wave as opposed to adiabatic.\\ 
\\ 
We explore the possibility that the MTS is correlated with the isotropic luminosity (L$_{iso}$). For the majority of the bursts, we use the luminosity values reported by Lu et al. (2012). We note here that we have only included those bursts for which the uncertainties in the derived $\Gamma$'s are given in the literature. There are a number of bursts reported by Lu et al. (2012) for which only limits on $\Gamma$ are reported -- these are not included in our sample. For the few Fermi GRBs, we have taken the luminosities found in the literature (Ghirlanda et al. 2012, Ackermann et al. 2010, Lu et. al. 2012, Maselli, et al. 2014, MacLachlan et al. 2013). The resulting plot of MTS (corrected for redshift) versus the isotropic luminosity is shown in Figure 4. As is clear from the figure, the MTS-Luminosity plot essentially exhibits the same features seen in the MTS-$\Gamma$ correlation (see Figure 1) i.e., a shallow/plateau phase (low luminosity) and a steeply declining phase (high luminosity). By in large, the low-luminosity region is dominated by bursts with small $\Gamma$'s and the high-luminosity bursts tend to have large $\Gamma$'s. We note that it is primarily the Fermi bursts with very high luminosities that particularly feature in the steep declining phase of the MTS-Luminosity plot while most of the Swift bursts tend to populate the plateau region. In passing, we note that the isotropic luminosities for the combined sample in our study (Lu et al. (2012) and the Fermi bursts) tightly follow the $\Gamma$-L$_{iso}$ correlation reported by Liang et al 2012 and Lu et al 2012. Indeed, Lu et al. (2012) construct a theoretical model and derive a jet luminosity powered by neutrino annihilation that is consistent with the correlation seen in the data. As MTS seems to correlates with the mean isotropic Luminosity (in the steep decline phase), we speculate that both of the newly found correlations i.e., MTS-$\Gamma$ and MTS-L$_{iso}$ are likely to be intrinsic in nature.\\
\\
However the recent study of Hascoet et al. (2014) suggests a word of caution; these authors point to possible observational selection effects in the extraction of $\Gamma$'s from the peak times associated with afterglow measurements. Lu et al. (2012) also considered such effects but concluded that the $\Gamma$-L$_{iso}$ correlation is intrinsic citing insignificant observational bias in the $\Gamma$ sample. With the question of selection bias and observational effects in mind, we have investigated, via a simulation, the possibility that the extracted MTS could suffer from such effects. We briefly describe the details of the simulation here: the GRB light curves were denoised with a hard-threshold wavelet process described by Donoho et. al in which wavelet coefficients smaller than some threshold are discarded. The remaining or the filtered coefficients are then used to reconstruct denoised versions of the light curves. These simulated lightcurves then serve as probability distribution functions (PDFs) to seed an event generator sampling from a Poisson distribution. The portion of the PDFs corresponding to the background level was held constant while the amplitude of the signal portion of the PDFs was scaled by various factors {25$\%$, 50$\%$, 100$\%$, and 200$\%$}, relative to the background level. We generated 1000 realizations of lightcurves at each of the scale factors. MTS values were extracted for the simulated lightcurves using the wavelet technique of MacLachlan et al 2012. The results are depicted in Fig. 4. The extracted MTS values indicate a dependence on the S/N ratio and as such should conservatively be considered upper limits for a given S/N ratio and for a given time scale especially for bursts that are noise dominated. However, very importantly, the main features of the plot i.e, the shallow/plateau region for low-luminosity bursts and the steep decaying phase for high-luminosity bursts, is still preserved thus indicating a possible underlying correlation.  The scatter does seem to increase the break between the two regions but the shift is within the noted uncertainties. We emphasize that the spectral lag also exhibits the same features as the MTS-$\Gamma$ correlation. Moreover, the extraction of the spectral lag does not suffer from the same level of sensitivity to the S/N as the MTS and is therefore likley to be a more robust indicator of the underlying correlation with the bulk Lorentz factor. We finish by noting that the question of possible observational selection effects warrants further investigation.\\
\\
\vspace{-0.5cm}
\section{Conclusions}

\hspace{-0.3cm} We summarize our main findings as follows:\\
\begin{itemize}

\item the MTS-$\Gamma$ plots exhibit a knee-like feature with a break around $\Gamma$ = 225$\pm$30. Beyond the break, the MTS decreases steeply with $\Gamma$ with a mean exponent of 4.05$\pm$0.64,

\item the shallow-plateau region of the MTS-$\Gamma$ correlation can be understood if one assumes that for low $\Gamma$'s, the radius of the emission site does not necessarily follow the often employed relation, $R\propto(\Gamma^{2}c\delta T)/(1+z)$, 

\item the declining phase of the MTS-$\Gamma$ correlation could be an opacity effect. A sufficiently large Lorentz factor is required to reduce the optical depth in order for high energy photons to escape without the creation of electron pairs. The extracted power-law exponent is $-4.05\pm0.64$, 

\item the MTS-L$_{iso}$ correlation exhibits the same features as the MTS-$\Gamma$ correlation. These two newly found correlations taken together (and barring significant selection effects) suggest an intrinsic connection between the prompt emission temporal variability (MTS) and the relativistic blast wave parameters, $\Gamma$ and L$_{iso}$, 

\item the spectral lag-$\Gamma$ plot exhibits a similar knee-like feature as the MTS-$\Gamma$ correlation,

\item we find an intriguing correlation between MTS, a prompt emission temporal variable, and the peak time, t$_p$, of the afterglow emission. The extracted power-law exponent (1.04$\pm$0.05) is consistent with the case of radiative dissipation of the blast wave although the adiabatic possibility is not ruled out by the data.
\end{itemize}

\hspace{-0.3cm}We conclude by noting that the MTS-$\Gamma$ correlation (in the steep decline phase) provides a method for extracting the bulk Lorentz factor for bursts directly from the observed temporal variability in the prompt emission. The MTS-t$_p$ correlation, if substantiated by more precise data, will provide a simple method for predicting the peak times for the optical emission.
\\
%
%
\begin{figure}
\epsscale{1.20}
\centering
\vspace{0.3 cm}
\plotone{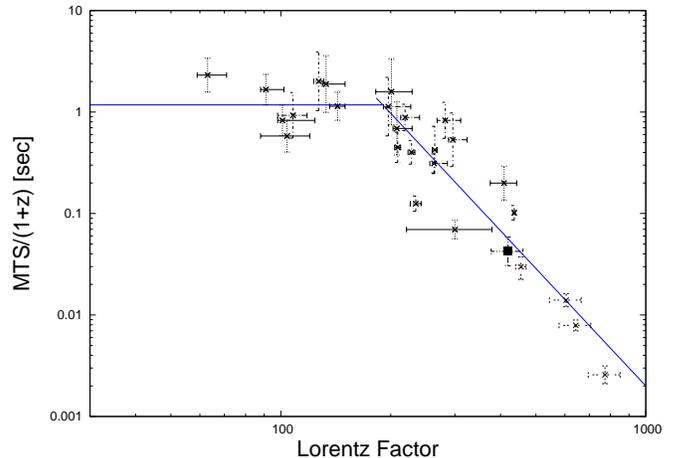}
\caption{MTS for the prompt emission vs Lorentz Factor (Lu et al. 2012). For GRB 080916C (shown as a filled square) the wind medium Lorentz Factor is plotted.}
\label{fig4}
\end{figure}

\begin{figure}
\epsscale{1.20}
\centering
\vspace{0.3 cm}
\plotone{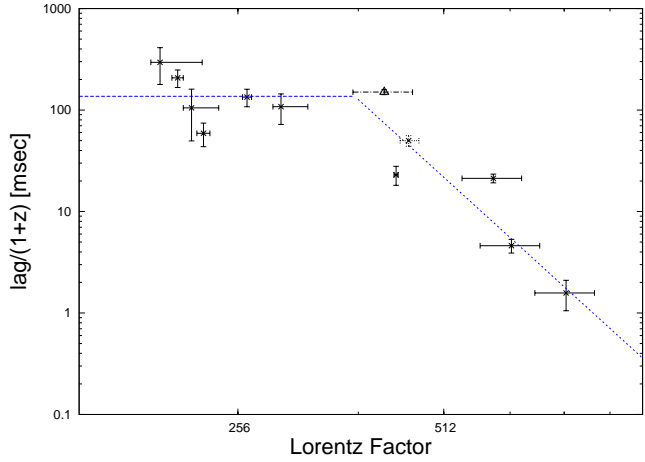}
\caption{Spectral Lag for the prompt emission vs Lorentz Factor (Lu et al. 2012).}
\label{fig4}
\end{figure}
\begin{figure}
\epsscale{1.20}
\centering
\vspace{0.3 cm}
\plotone{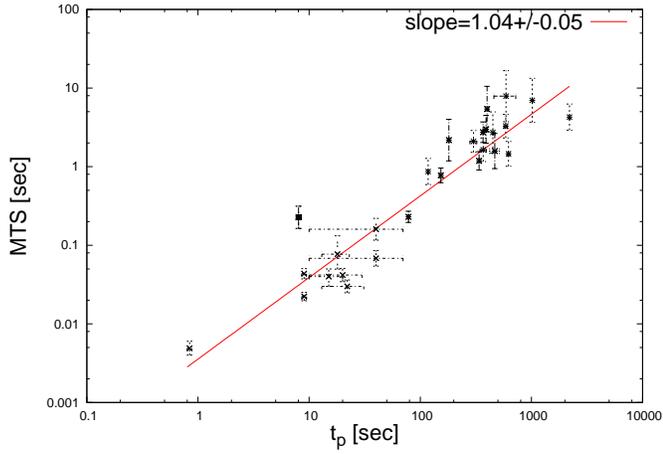}
\caption{MTS for the prompt emission vs. peak times (t$_p$) for the optical afterglow. For the Fermi GRBs afterglow peak times are taken from Ghisellini et al. (2010) and Ackerman et al. (2013). GRB 080916C is shown as a filled square.}
\label{fig2}
\end{figure}
\begin{figure}
\epsscale{1.20}
\centering
\vspace{0.3 cm}
\plotone{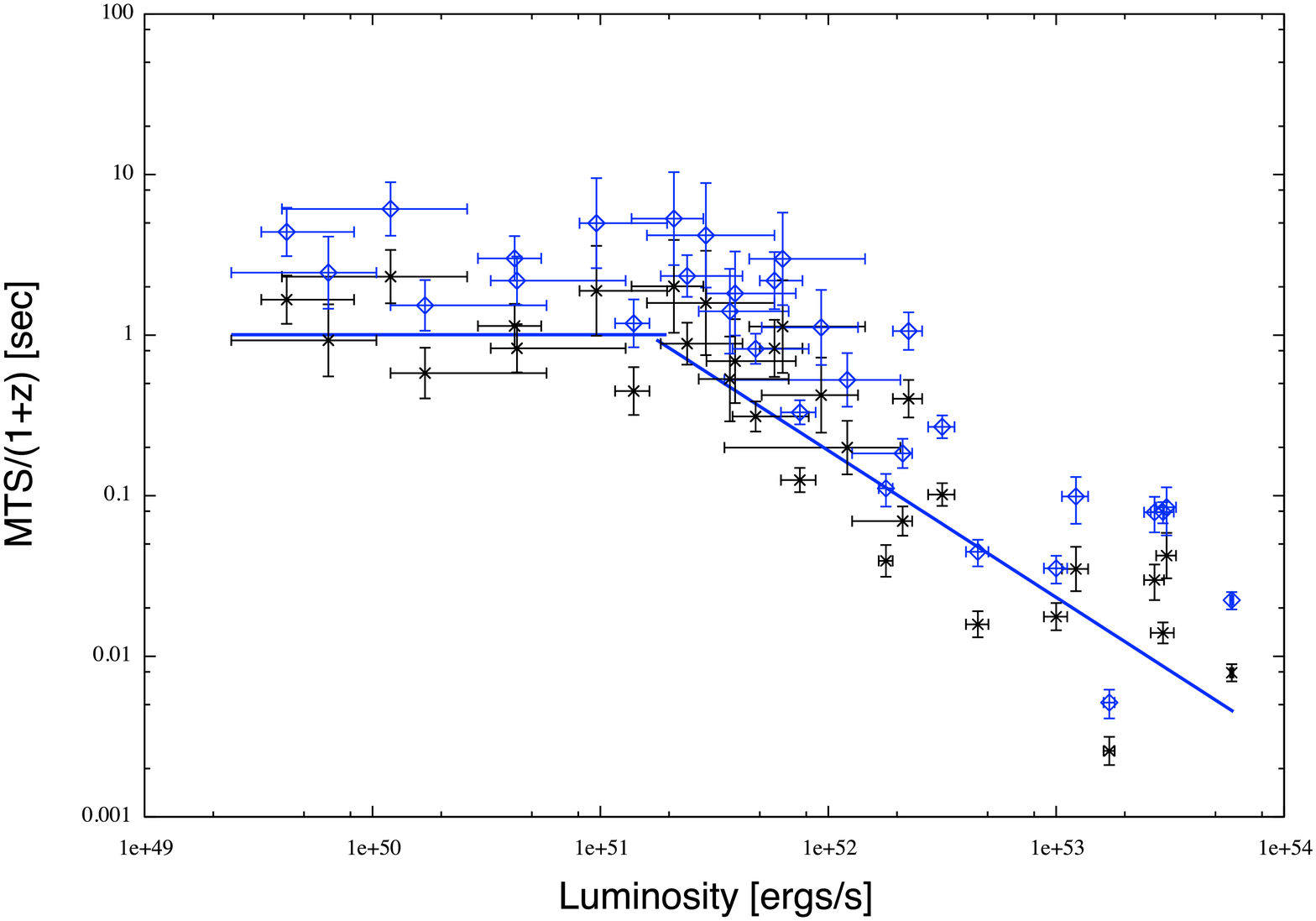}
\caption{MTS (corrected for redshift) for the prompt emission vs. isotropic Luminosity. Blue points are for simulated light curves.}
\label{fig2}
\end{figure}

\section{Acknowledgments}
\hspace{-0.3cm} E.S. acknowledges financial support from the George Washington University (W. J. Briscoe, GWINS) and The Science Academy (Bilim Akademisi, Turkey) under the BAGEP program.

\newpage
\begin{table*}[ht] 
\caption{Minimum variability times, afterglow peak times and bulk Lorentz factors for selected GRBs} 
\centering 
\begin{tabular}{cccccccc} 
\hline\hline 
GRB Name & $\tau$ [sec] & $\delta\tau{^-}$ [sec] & $\delta\tau{^+}$ [sec] & $\Gamma$\footnote{Ghirlanda et al. 2012} &  $\Gamma$
\footnote{Lu et al. 2012} & t$_p$  \footnote{Liang et al. 2010} \\ 
\hline 
GRB 050730 & 7.88 & 0.88 & 17.61 &- & 201$^{+28}_{-19}$ &  590.7 $\pm$ 131.5 \\ 
GRB 050820A & 2.99 & 0.65 & 3.39 & 142 & 282$^{+29}_{-14}$ &   391.0 $\pm$ 16.7 \\
GRB 050922C & 0.56 & 0.22 & 0.36 & 138 & - & -\\
GRB 060210 & 2.07 & 0.35 & 3.04 & 133 & 264$\pm$4  & - \\
GRB 060418 & 0.78 & 0.14 & 0.59 & 137 & 263$^{+23}_{-7}$ & 153.3 $\pm$ 3.3\\
GRB 060605 & 5.40 & 0.72 & 10.20 & 101 & 197$^{+30}_{-6}$ &  399.1 $\pm$ 13.0  \\
GRB 060607A & 2.18 & 0.32 & 3.67 & 153 & 296$^{+28}_{-8}$&  180.9 $\pm$ 2.4 \\
GRB 060904B & 1.58  & 0.28 & 2.22 & 50 &108$\pm$10 & 467.9 $\pm$ 48.4\\
GRB 061007 & 0.23 & 0.08 & 0.15 & 215 & 436$\pm$3 & 78.3 $\pm$ 0.4 \\
GRB 061121 & 0.29 & 0.10 & 0.20 & 88 & 175$\pm$2 & - \\
GRB 070110 & 6.75 & 0.89 & 12.78 & 64 & 127$\pm$4 & -\\
GRB 070318 & 2.09 & 0.55 & 1.98 & - & 143$\pm$7 &301.0 $\pm$ 21.3\\
GRB 070411 & 2.72  & 0.41 & 4.54 & - & 208$^{+21}_{-5}$ &  450.1 $\pm$ 5.0 \\
GRB 070419A & 3.27  & 0.82 & 3.27  & - & 91$^{+11}_{-3}$ &  587.0 $\pm$ 20.9\\
GRB 071010A& 1.64 & 1.64 & 0.41 & - & 101$^{+23}_{-3}$& 368.2 $\pm$ 24.4\\
GRB 071010B& 0.873& 0.22& 0.87& 105 & 209 $\pm$ 4&  -\\
GRB 071031 & 6.96 & 0.96 & 12.65 & - & 133$^{+17}_{-3}$ &  1018.6 $\pm$1.6\\
GRB 080319B & 0.077 & 0.027 & 0.055 &  - & - & 18$\pm$5\footnote{Racusin et al. 2008} \\
GRB 080319C & 1.19 & 0.35 & 1.01 & 109 & 228$\pm$5 &338.3 $\pm$ 5.6\\
GRB 080330 & 1.46 & 0.35  & 1.51 & - & 104$^{30}_{-2}$&621.9 $\pm$ 17.0  \\
GRB 080710 &  4.27 & 0.99 & 4.60 & - & 63$^{+8}_{-4}$&  2200.9 $\pm$ 4.1\\
GRB 080804 & 2.76 & 0.57  & 3.33 & 157 & - & -\\
GRB 080810 & 0.87 & 0.20 & 0.94 & 214 & 409$\pm$34&  117.6 $\pm$ 1.1\\
GRB 080916C & 0.2266\footnote{MacLachlan et al. 2013} & 0.0630 & 0.0872 & 419 & - & 8.03\footnote{Ghisellini et al. 2010} \\
GRB 081203A & 2.74 & 0.75 & 2.49 & 121 & 219$^{+21}_{-6}$ & 367.1 $\pm$ 0.8 \\
GRB 090102 & 2.11 & 0.34 & 3.23 & 221 & - & - \\
GRB 090323 & 0.1598$^{e}$ & 0.0436 &  0.0599 & - & - & 40\footnote{Ackerman et al. 2013}$\pm$30 \\
GRB 090328 & 0.0682$^{e}$ & 0.0139 &  0.0175 & - & - & 40$^{g}$$\pm$30 \\
GRB 090424 & 0.11 & 0.04 & 0.08 & - & 300$\pm$79 &  -\\
GRB 090510 & 0.0049$^{e}$ & 0.0009 & 0.0011 & 773 & - & 0.84$^{f}$ \\
GRB 090618 & 0.38 & 0.13  & 0.28 & 158 & - & -\\
GRB 090812 & 1.23 & 0.36 & 1.05 & 253& - & -\\
GRB 090902B & 0.0223$^{e}$ & 0.0029 & 0.0026 & 643 & - & 9.03$^{f}$ \\
GRB 090926A & 0.0435$^{e}$ & 0.0061& 0.0070 & 605 & - & 9.01$^{f}$ \\ 
GRB 091003 & 0.0300 & 0.0051 & 0.0062& - & - & 22$^{g}$$\pm$9 \\
GRB 091024 & 4.31 & 0.85 & 5.46 & 59 & - & -\\
GRB 091029 & 3.02 & 0.86 & 2.65 & 111  & - & -\\
GRB 100414 & 0.0418 & 0.0074 & 0.0090& - & - & 20$^{g}$$\pm$10 \\
GRB 100621A & 0.92 & 0.29 & 0.71 & 26  & - & -\\
GRB 100728B & 1.81 & 0.27 & 2.98 & 188 & -& -\\
GRB 100906A & 0.75  & 0.24 & 0.59 & 186 & - & -\\
GRB 110205A & 0.60 & 0.09 & 1.02 & 89 & -& -\\
GRB 110213A & 1.11 & 0.36  & 0.85 & 113 & - & -\\
GRB 130427A & 0.04\footnote{Ackerman et al. 2014} & 0.01 & 0.01 & 455$^{h}$$\pm$15 & - & 15$\pm$5 \\ 
\hline 
\\
\\
 \\
 \\
\\
\end{tabular} 
\label{table:nonlin} 
\end{table*}

\end{document}